\documentclass[a4paper]{article}
\pdfoutput=1
\usepackage{INTERSPEECH2021}
\usepackage{cite}
\usepackage{bm}
\usepackage{multirow}
\usepackage{color}
\usepackage{mathrsfs}
\usepackage{amsmath}

\title{CNN-based Discriminative Training for Domain Compensation in Acoustic Event Detection with Frame-wise Classifier}
\name{Tiantian Tang$^{1}$, Xinyuan Zhou$^{1}$, Yanhua Long$^{*1}$\thanks{Yanhua Long is the corresponding author. The work is supported by the National Natural Science Foundation of China (No.62071302).}, Yijie Li$^2$, Jiaen Liang$^2$}

\address{
  $^1$Shanghai Normal University, Shanghai, China\\
  $^2$Unisound AI Technology Co., Ltd., Beijing, China}
  \email{1000479042@smail.shnu.edu.cn,yanhua@shnu.edu.cn}

\begin{document}

\maketitle
\begin{abstract}


Domain mismatch is a noteworthy issue in acoustic event detection tasks,
as the target domain data is difficult to access in most real applications.
In this study, we propose a novel CNN-based discriminative training
framework as a domain compensation method to handle this issue.
It uses a parallel CNN-based discriminator to learn a pair of high-level intermediate
acoustic representations. Together with a binary discriminative
loss, the discriminators are forced to
maximally exploit the discrimination of heterogeneous acoustic information
in each audio clip with target events, which results in a robust paired representations
that can well discriminate the target events and background/domain
variations separately. Moreover, to better learn
the transient characteristics of target events, a frame-wise classifier
is designed to perform
the final classification. In addition, a two-stage training with
the CNN-based discriminator initialization is further
proposed to enhance the system training. All experiments are
performed on the DCASE 2018 Task3 datasets. Results show that
our proposal significantly outperforms the official baseline on cross-domain
conditions in AUC by relative $1.8-12.1$\% without any performance degradation
on in-domain evaluation conditions.


%
%
%
%

\end{abstract}

\section{Introduction}
\label{sec:intro}

%

Acoustic event detection (AED) refers to the task of detecting whether interested
target events occur in audios such as running water, cough, meow, etc. With the
launch of the Detection and Classification of Acoustic Scenes and Events (DCASE)
challenges from 2013\cite{Stowell2015}, a set of AED-related tasks are provided for research and
progress comparison of state-of-the-art techniques. The bird audio detection (BAD)
\cite{Stowell_2018badchj} task is the DCASE 2018 Task3 that aims to detect
the presence/absence of bird sound in audio clips under variety bird species,
recording and background conditions. To solve this task well,
the approaches are required to inherently generalize across conditions
or can be self-adapted to new datasets, because there is big domain mismatch between
training and evaluation sets. In this study, we also focus on the domain mismatch issue
for BAD task, because the source and target domain mismatch is a common problem
in most AED tasks\cite{wei2020crnn,Fonseca2019}, and it typically results in a severe performance degradation
in practical applications \cite{ben2010theory}.

In the literature, only few previous works have been proposed to improve the
domain robustness of BAD systems. Such as in \cite{lostanlen2019robust}, authors
applied a per-channel energy normalization to alleviate the
outdoor acoustic environment distortions.
In \cite{berger2018bird}, authors used the wasserstein distance guided
representation learning \cite{shen2018wasserstein} to incorporate
the domain knowledge during model training. And works in \cite{liaqatdomain} applied
the CORrelation ALignment  \cite{sun2016return} to
minimize the domain shift by aligning the second-order statistics
of source and target distributions. There are also
some domain compensation or adaptation
methods are proposed for other acoustic processing tasks
\cite{asami2017domain,mun2019domain,hubeika2008discriminative,8683616,
duroselle2020metric,mirsamadi2017multi}.
For example, in acoustic scene classification task,
a spectrum correction \cite{Komider2019} was proposed
to corrected the mismatched front-end by adjusting the varying
frequency response of different recording devices.
\cite{hu2020relational} proposed a neural label embedding together
with a relational teacher-student learning to perform the device
adaptation. And in \cite{Gharib2018}, the unsupervised adversarial
learning was used to leverage an extra domain discriminator for
device adaptation and it was further generalized for AED tasks in \cite{wei2020crnn}.


Unlike previous domain adaptation methods, in this study, we
deal with the domain mismatch in BAD tasks by
proposing a novel discriminative training framework with two
CNN-based discriminators, where each input audio clip is transformed
into a pair of discriminative high-level acoustic representations
before feeding them to the back-end binary classifier.
This is motivated by the intuition that extracting the high-level
representation using a standard neural network such as
CNN or LSTM optimized only by the final task-dependent loss
might not be the best choice, as it may tend to be trapped
in local optima and fail to extract the fine-grained heterogeneous
acoustic information between targets and interferences we need.
Therefore, we wonder if it is possible to learn two discriminative
representations from each audio clip instead of one
to enhance the domain robustness of AED systems. In the BAD task,
we design a two-stage training strategy with a binary discriminative
loss to force the CNN-based discriminators
to learn the acoustic discrimination between bird calls and background interferences
separately. The resulted paired representations are then feed into
a specially designed frame-wise classifier to further capture the
transient characteristics of bird calls.
All experiments are performed on the DCASE 2018 Task3 datasets. Compared
with the official baseline system, results show that
our proposed framework can achieve significant performance improvements
on cross-domain test conditions without degrading performance
of in-domain test conditions.

\section{Proposed Method}
\label{sec:pro}

\subsection{Architecture}
\label{subsec:arc}

\begin{figure*}[t]
  \centering
  \includegraphics[width=12cm]{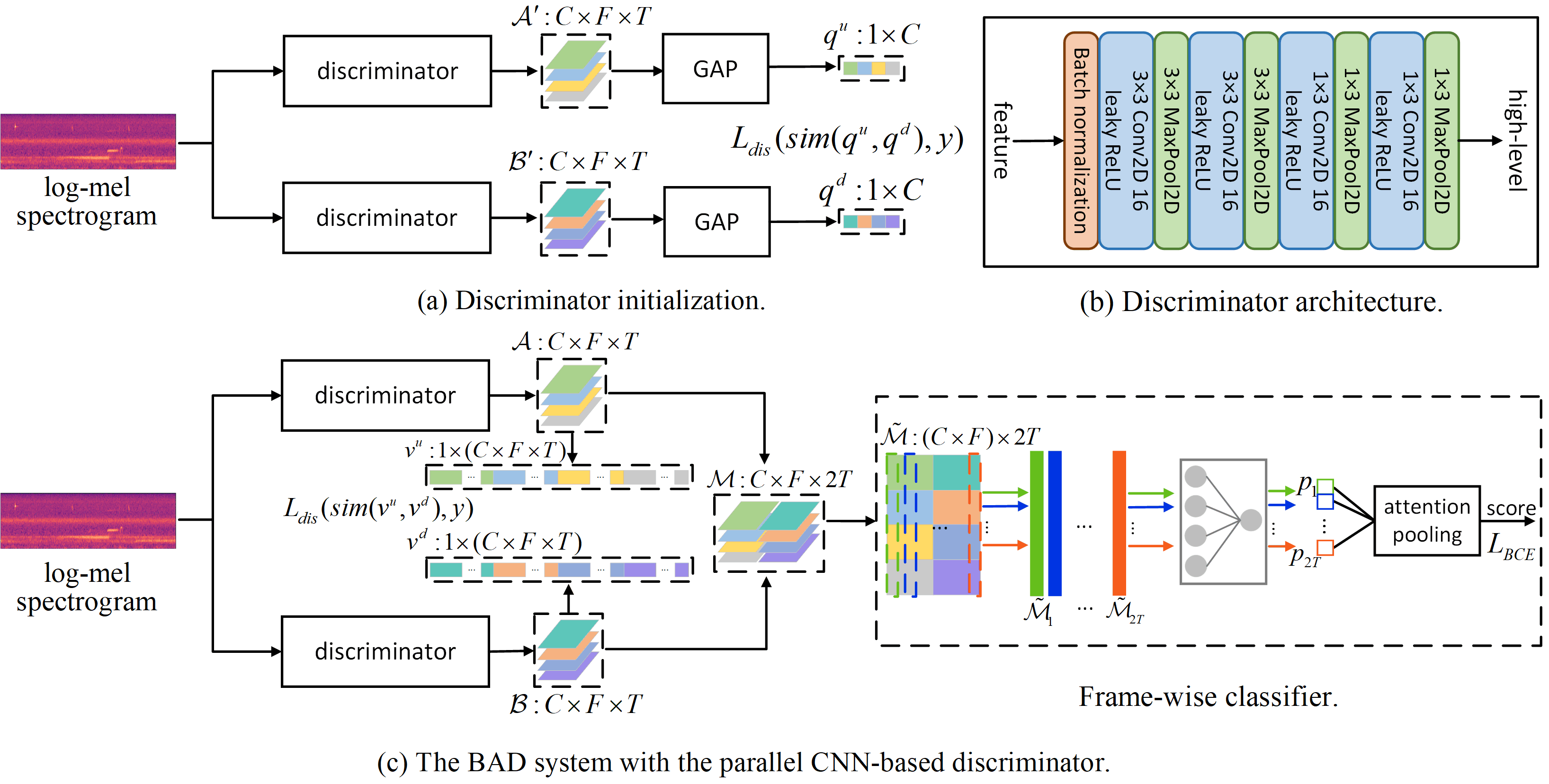}
  \caption{Framework of the proposed CNN-based discriminative training with frame-wise classifier.}
  \label{fig:pro}
\end{figure*}


Fig. \ref{fig:pro} gives an overview of our proposed method.
As in block (c), the BAD system consists of two parts:
a parallel CNN-based discriminator and followed by a frame-wise binary classifier.
The discriminators are with the same structure as shown in block (b).
Given a labeled dataset $\mathcal{D} = \{ ({x_i},{y_i})\} _{i = 1}^N$
with $N$ samples, where $x_i$ is the input audio clip and
$y_i \in \{0,1\}$ is a label to indicate the absence/presence of any bird calls
within that clip. Each input $x$ is first transformed to a
log-mel spectrogram ${X} \in {\mathbb{R}^{1 \times F \times T}}$, where $F$ and
$T$ are the number of mel-frequency bins and frames respectively. By taking
$X$ as input feature, the parallel discriminator $d_{cnn}(\cdot,\theta)$
with $C$-channel output layers are employed to obtain a pair of intermediate discriminative
representations ${\mathcal{A}} \in {\mathbb{R}^{C \times F \times T}}$ and ${\mathcal{B}} \in {\mathbb{R}^{C \times F \times T}}$,
\begin{equation}
\label{eq1}
 {\mathcal{A}} = d_{\rm cnn}({X},{\theta _u})
\end{equation}
\begin{equation}
\label{eq2}
{\mathcal{B}} = d_{\rm cnn}({{X},\theta _d})
\end{equation}
where $\theta_u$ and $\theta_d$ denote the CNN parameters of the upper
and down discriminators in Fig. \ref{fig:pro} (c) respectively.
Then we concatenate ${\mathcal{A}}$ and ${\mathcal{B}}$ into a tensor
${\mathcal{M}} \in {\mathbb{R}^{C \times F \times 2T}}$ as
\begin{equation}
\label{eq3}
 \mathcal{M} = [{\mathcal{A}} , {\mathcal{B}}]
\end{equation}
Next, we reshape $\mathcal{M}$ into a $(C \times F) \times 2T$
matrix $\mathcal{\tilde{M}}$ as input of the frame-wise
classifier $\mathcal{G}(\cdot)$ to perform the binary classification.
Details of the parallel discriminator, the frame-wise binary classifier
and system training strategy are presented in the following subsections.


\subsection{CNN-based Discriminator}
\label{subsec:cnn}



Motivated by the fine-grained structure extraction \cite{chang2020devil} and
domain-invariant representation learning \cite{bousmalis2016domain} for
image classification, here we investigate to
use two same structure CNN networks as a parallel discriminator
that shown in Fig.\ref{fig:pro} (c) to
extract a pair of intermediate discriminative representations
to enhance the bird calls detection system. We aim to mine the
intermediate feature discrimination of heterogeneous acoustic information
that embedded in each input audio clip, such as the
target events (bird calls) and variety background interferences in a given clip
of BAD task.

To force the designed parallel discriminator to well learn the heterogeneous
acoustic characteristic separately,
we introduce a novel binary discriminative loss as a training criteria to
trained the parallel discriminator simultaneously as below:
\begin{equation}
\label{eq4}
{L_{\rm dis}(s,y)} =  - [y\log (1 - {\mathcal{H}}({s})) + \lambda(1 - y)\log (\mathcal{H}({s}))]
\end{equation}
where $y\in \{0,1\}$ denotes the ground-truth label of input clip,
$\lambda\in[0,1]$ is a tuning parameter to balance the loss contribution of positive and negative
training data. ${\mathcal{H}}(\cdot)$ is the rectified linear unit (ReLU) to ensure that the $s$ is non-negative. $s$ is a cosine similarity measure that defined as :
\begin{equation}
\label{eq5}
{s} = {\rm sim}(\mathcal{T}({\mathcal{A}} ),\mathcal{T}({\mathcal{B}} )) = \frac{{\mathcal{T}({\mathcal{A}} ) \cdot \mathcal{T}({\mathcal{B}} )}}
{{\left\| {\mathcal{T}({\mathcal{A}} )} \right\|\left\| {\mathcal{T}({\mathcal{B}} )} \right\|}}
\end{equation}
where $\mathcal{T}(\cdot)$ is the flatten operation that transforms a
tensor into a vector,  as show in Fig. \ref{fig:pro} (c),
the $\mathcal{T}({\mathcal{A}} ),\mathcal{T}({\mathcal{B}} )$ transforms
the $\mathcal{A}, \mathcal{B}$ into vector ${v}^u, {v}^d$ respectively.

%
%
%


Based on Eq.(\ref{eq4}), we minimize the ${L_{\rm dis}(s,y)}$
to achieve our goal. It means that when the input
clip is a positive sample ($y=1$) with heterogeneous acoustic information, i.e., the
clip is a mixture signal that contains both bird calls and background noises,
then only the 1st part $y\log (1 - {\mathcal{H}}({s}))$ contributes to
${L_{\rm dis}(s,y)}$ and maximum ${\mathcal{H}}({s}) = 0$ (
${\mathcal{A}}$ and ${\mathcal{B}}$ are totally different, one learns background
characteristics, while the other emphasizes the bird calls).
When the input is a negative sample ($y=0$, only background sounds),
only the 2nd part of Eq.(\ref{eq4}) contributes to the loss,
then the discriminators do not differentiate their outputs, because minimizing
${L_{\rm dis}(s,y)}$  leads to a maximum value of ${\mathcal{H}}({s}) = 1$, it
results similar paired ${\mathcal{A}}$ and ${\mathcal{B}}$.
That's to say, ${L_{\rm dis}(s,y)}$  only discriminates the heterogeneous
information in positive samples to highlight the acoustic characteristics
of target events. These target representations should be more robust to
domain variation, they can be taken as domain-invariant intermediate features
because they capture the acoustic properties of bird calls more explicitly.

In most real applications as our BAD task, the positive samples
are always with background sounds. Therefore, we think that if the BAD
system is trained on a source domain, the robust target representations
learned from the parallel CNN-based discriminator
can leverage a better model generalization to the target domain bird call detection.
That's to say, the discriminators play a domain compensation role in the whole BAD system.


\subsection{Frame-wise Binary Classifier}
\label{subsec:fra}

Different from other bioacoustics signals, the bird calls are normally
short, their spectrograms have strong transient characteristics.
Instead of using the conventional classifier
of official baseline \cite{grill2017two} that accepts the
whole flattened CNN feature maps as one input, here we propose to use
a frame-wise classifier to learn the
transient bird chirping characteristics. As shown in the
last block of Fig.\ref{fig:pro}, each column
${\tilde {\mathcal{M}}_j} \in {\mathbb{R}^{C \times F}}$, $j = 1,2, \dots ,2T$,
is taken as $j$-th frame, then each ${\tilde {\mathcal{M}}_j}$
is learned independently by a two-layers feed-forward neural network (FFN)
followed with a sigmoid activation to achieve a prediction score $p_{j}$.
All ${\tilde {\mathcal{M}}_j}$ share the same FFN parameters.
Finally, all the $2T$ prediction scores are further attention-weighted
by the attention pooling \cite{xu2018large,kong2018audio}
to automatically control their contribution for decision making.
The final score $p$ is computed as:
\begin{equation}
\label{eq7}
{p} = ({\sum\limits_j {{p_{j}w_j}} })/{{\sum\limits_j {w_j} }}
\end{equation}
where $w_j$ is the learnable weight for each $p_{j}$. Details of
the attention pooling can be found in \cite{wang2019comparison}.

\subsection{Two-stage Training Strategy}
\label{subsec:tra}

In intuition, the proposed BAD system in Fig.\ref{fig:pro}(c)
should be trained in one-stage using a combination loss that defined as:
\begin{equation}
\label{eq9}
L_{\rm total} = {L_{\rm dis}}({\rm sim}({v^u},{v^d}),y) + {L_{\rm BCE}({{\hat y},y})}
\end{equation}
where ${\hat y}$ is the final prediction score $p$, ${L_{\rm dis}}(\cdot)$ is the binary discriminative loss defined in Eq.(\ref{eq4}), and ${L_{\rm BCE}(\cdot)}$ is the
tradition binary cross entropy (BCE) loss as in \cite{lasseck2018acoustic,himawan20183d}.
${v^u},{v^d}$ are the flattened representations used in Eq.(\ref{eq5}).

However, from our extensive tryout experiments, we find that it's better
to use a two-stage training strategy with the parallel CNN-based
discriminator initialization. In \verb"stage 1", as shown in Fig.\ref{fig:pro}(a),
we only train the parallel discriminator using a binary discriminative loss defined as,
\begin{equation}
\label{eq10}
{L_{\rm pre}} = {L_{\rm dis}}({\rm sim}({{q}^u},{{q}^d}),y)
\end{equation}
where ${q^u} = {\rm GAP}({\mathcal{A}} ')$,  ${q^d} = {\rm GAP}({\mathcal{B}} ')$ , the
${\mathcal{A}} ', {\mathcal{B}} '$ and the size of ${q^u},{q^d}$ are illustrated in
Fig.\ref{fig:pro}(a).  The GAP  denotes using the global average pooling \cite{lin2013network} to map each channel representation into a average one.
It is different from the flatten that used in one-stage training loss.

Based on the well pre-trained discriminators, in  \verb"stage 2",
the whole system is then trained using the above combination loss $L_{\rm total}$,
but with the discriminators are initialized by the pre-trained
CNN parameters in  \verb"stage 1". We speculate that an effective
initialization may avoid local optima and provide a good guidance to enhance
the whole model training, because the pre-trained discriminators  can
provide a stable and discriminative perception to the frame-wise classifier.

\section{Experimental Setup}
\label{sec:exp}

\subsection{Dataset}
\label{subsec:data}

The DCASE 2018 Task 3 (bird audio detection) provides 3 separate
labeled development and 3 evaluation datasets,
each recorded under different conditions. As the ground-truth
of evaluation set is not released publicly. Only the development
sets are used in our work. The datasets have different balances of
positive/negative cases, different bird species and a wide-domain coverage of
background sounds and recording equipments. Each audio clip is 10s-length
and sampled at 44.1kHz.

Specifically, three development sets are the
``freefield1010" (ff1010bird), the ``warblrb10k" and the
 ``BirdVox-DCASE-20k" (BirdVox-20k).
The ff1010bird contains 7,690 excerpts from field recordings
around the world with a diverse location and environments.
The warblrb10k contains 8,000 smartphone audio recordings from around the UK,
the audio covers a wide distribution of UK locations and environments, and it
includes weather/traffic noise, human speech and even human bird imitations.
The BirdVox-20k consists of 20,000 audio clips that collected from remote
monitoring units placed near Ithaca, NY, USA during the autumn of 2015.
Compared with ff1010bird and BirdVox-20k, the warblrb10k contains
much more diverse background acoustics.
Instead of using the experimental procedure recommended
by DCASE challenge to achieve one general model,
our goal is to examine the model generalization ability
for cross-domain evaluation tasks, so we construct our own BAD tasks using
the provided development sets, for each of the above mentioned
dataset, we select $60$\%, $20$\%, $20$\% audio clips for training,
validation and test respectively.

\subsection{Features and Models}
\label{subsec:fm}

Each clip is down-sampled to $22.05$ kHz and then
divided into $46$ ms frames using hanning window
with a hop size of 14 ms.  80-dimensional
log-mel filter banks extracted across a frequency range
from 50 to 11 kHz are used as input features for both
the baseline and our method. The official ``Area Under the Curve (AUC) of
Receiver Operating Characteristic curve (ROC)" \cite{dcasewebsite}
is used to evaluate the system performances.





The official baseline of DCASE 2018 BAD challenge  \cite{grill2017two} is taken
as our baseline. Its also a CNN-based encoder-classifier structure. The CNN-based encoder
is the same as our discriminator as shown in block (b) of Fig.\ref{fig:pro}.
This encoder is then followed by three dense (fully-connected) layers with $256$, $32$ and $1$ unit(s) as a binary classifier. Each convolution and dense layer use the
leaky rectifier nonlinearity as their activation function except for the sigmoid output layer.

Different from baseline, our model Fig.\ref{fig:pro} (c) uses
two same structure CNN-based encoders as a parallel discriminator,
but the followed frame-wise classifier only has two dense layers with
$32$ and $1$ unit(s). Besides using the frame-wise classifier,
as the baseline model, we also investigate to use the conventional dense
layers as the classifier (F-C) to learn the directly flattened
discriminator outputs $\mathcal{\tilde{M}}$.
As the flattened vector dimension is too large (2816) than that in the baseline,
a four dense layers ($512$, $256$, $32$ and $1$ unit(s)) instead of three
is used in the F-C to achieve a better results.
The Adam Optimizer \cite{kingma2014adam} with a learning rate
of ${10^{ - 4}}$ is used for both the baseline and one-stage training.
The two-stage training  uses ${10^{ - 3}}$ as initial
learning rate, and then gradually decaying to ${10^{ - 5}}$ in \verb"stage 1",
then fixed to ${10^{ - 4}}$ in \verb"stage 2". $200$ epochs are used in each stage.


%
%

%
%
%
%
%
%



\section{Results}
\label{sec:res}


\subsection{Results with one-stage training}
\label{subsec:epm}

Table \ref{tab:rpm} shows the performance comparison
using one-stage training.
Two training-test tasks are constructed to evaluate the effectiveness
of the proposed methods. One is using ``BirdVox-20k" as the
training set while the other is using ``warblrb10k" to train the model.
Both of them are tested on the same three subset of ``BirdVox-20k, warblrb10k
and ff1010bird", there is no clip overlap between training and test data.



\begin{table}[!ht]
  \caption{Results (AUC\%) of the proposed model with F-C or
  frame-wise (F-W) classifier using one-stage training strategy.
$\lambda=0.1$ and 1.0 in Eq.(\ref{eq4}) achieve the best results for both F-C and F-W that shown in the 1st and 2nd blocks respectively.}
  \label{tab:rpm}
  \centering
  \scalebox{0.9}{
  \begin{tabular}{l | l | l r r}
    \toprule
    \textbf{Train set} &\textbf{Test set} &\textbf{Baseline} &\textbf{F-C} &\textbf{F-W}   \\
    \midrule
    \multirow{3}{*}{\shortstack{BirdVox-20k}}
    &BirdVox-20k &94.62 &94.57 &93.63      \\
    &warblrb10k &62.57 &69.98 &68.96           \\
    &ff1010bird &75.11 &79.42 &79.61          \\
    \midrule
    \midrule
    \multirow{3}{*}{\shortstack{warblrb10k}}
    &warblrb10k &94.29 &94.39 &94.66            \\
    &BirdVox-20k &64.33 &65.74 &68.37     \\
    &ff1010bird &85.22 &82.98 &86.47           \\
    \bottomrule
  \end{tabular}}
\end{table}

From the baseline results of Table \ref{tab:rpm}, it's clear that
there are big performance gaps between in-domain and cross-domain tasks.
Results on the in-domain test sets are much
better than those on cross-domain test sets. By comparing
the F-C and baseline results, we see significant
AUC improvements on the cross-domain test tasks, such as
when we train the model on ``BirdVox-20k", there are relative
11.8\% and 5.7\% improvements on the ``warblrb10k" and ``ff1010bird"
respectively. In the 2nd block of Table \ref{tab:rpm}, we only achieve
limited gains (relative 2.2\% on ``BirdVox-20k")
or even a little bit worse (relative 2.6\% on ``warblrb10k") results when the
model is trained on a very wide-domain acoustic coverage dataset
``warblrb10k".
These improvements indicate that
the proposed CNN-based discriminator
is very effective to enhance the cross-domain performances when 
the model is trained on  ``BirdVox-20k" that with no richness background acoustics.
Because: 1) both the baseline and F-C system are with the same type of
classifiers; 2) as shown in section \ref{subsec:data},
the  ``warblrb10k" is very diverse that contains a rich acoustic environment while ``BirdVox-20k" is recorded from a fixed place with remote monitoring units.



Interestingly, by comparing the results in last two columns of
Table \ref{tab:rpm}, we see that under one-stage training strategy,
almost no improvements can be found when the model is trained on ``BirdVox-20k" ,
however, the proposed frame-wise classifier achieves
around absolute 2.6-3.5\% AUC improvements over the F-C on the cross-domain
test sets when the model is trained on ``warblrb10k". 
In addition, it's clear to see that there is almost no
performance change on the in-domain test set results, either
for the ``BirdVox-20k" or  ``warblrb10k" in-domain tasks,
it indicates that both of the proposed CNN-based discriminative training
and the frame-wise classifier are effective to improve the cross-domain
BAD performances without worsening any in-domain performances.

\subsection{Results with two-stage training}
\label{subsec:rwt}

\begin{table}[!ht]
  \caption{Comparison (in AUC\%) of training strategy with and without discriminator initialization.
  F-W is the one-stage training, TS-fla and TS-GAP represent two-stage training strategy
  using the $L_{\rm pre}$ with flatten and GAP operation respectively. $\lambda=0.3$ and 0.1 in Eq.(\ref{eq4}) achieve the best results for both TS-fla and TS-GAP that shown in the 1st and 2nd blocks respectively.}
  \label{tab:pre}
  \centering
  \scalebox{0.9}{
  \begin{tabular}{l|l|lll}
    \toprule
    \textbf{Train set} &\textbf{Test set} &\textbf{F-W} &\textbf{TS-fla} &\textbf{TS-GAP} \\
    \midrule
    \multirow{3}{*}{\shortstack{BirdVox-20k}}
    &BirdVox-20k &93.63 &93.54 &94.29      \\
    &warblrb10k &68.96 &66.28 &70.13            \\
    &ff1010bird &79.61 &79.13 &82.52            \\
    \midrule
    \midrule
    \multirow{3}{*}{\shortstack{warblrb10k}}
    &warblrb10k &94.66 &94.23 &94.86            \\
    &BirdVox-20k &68.37 &63.64 &68.61     \\
    &ff1010bird  &86.47 &86.48 &86.73           \\
    \bottomrule
  \end{tabular}}
\end{table}

\vspace*{-0.2cm}
Table \ref{tab:pre} shows the results of the proposed method with
two-stage training using different CNN-based discriminator initialization. 
Comparing the results of TS-GAP with TS-fla, we see that pre-training
the discriminators using $L_{\rm pre}$ with GAP achieves much better
results than using flatten operation. Performances of systems
using $L_{\rm pre}$ with flatten as initialization are even worse
than the ones from one-stage training strategy.
This may due to the fact that global average pooling as a structural regularizer sums out the spatial information, which is less prone to overfitting than traditional flatten operation \cite{lin2013network}.
Furthermore, when comparing the TS-GAP with F-W,
it's clear that the performances from two-stage training
is slightly better than the ones from one-stage training on all in-domain
and cross-domain tasks. However, the gains shown in the 1st block of Table \ref{tab:pre}
are much larger than the ones shown in the 2nd block. This phenomenon
is consistent with the observation from Table \ref{tab:rpm}.
Finally, by comparing the AUCs of the TS-GAP and the baseline,
our proposed method can bring relative 12.1\%, 9.9\% and 6.7\%, 1.8\%
AUC improvements over baseline on the ``BirdVox-20k"
and ``warblrb10k" based cross-domain tasks, respectively.
These gains also indicate that the proposed discriminative training
is more effective when there is large background domain mismatch between
training and test data.


\subsection{Visualization}
\label{vis}

\begin{figure}[!ht]
  \centering
  \includegraphics[width=6cm]{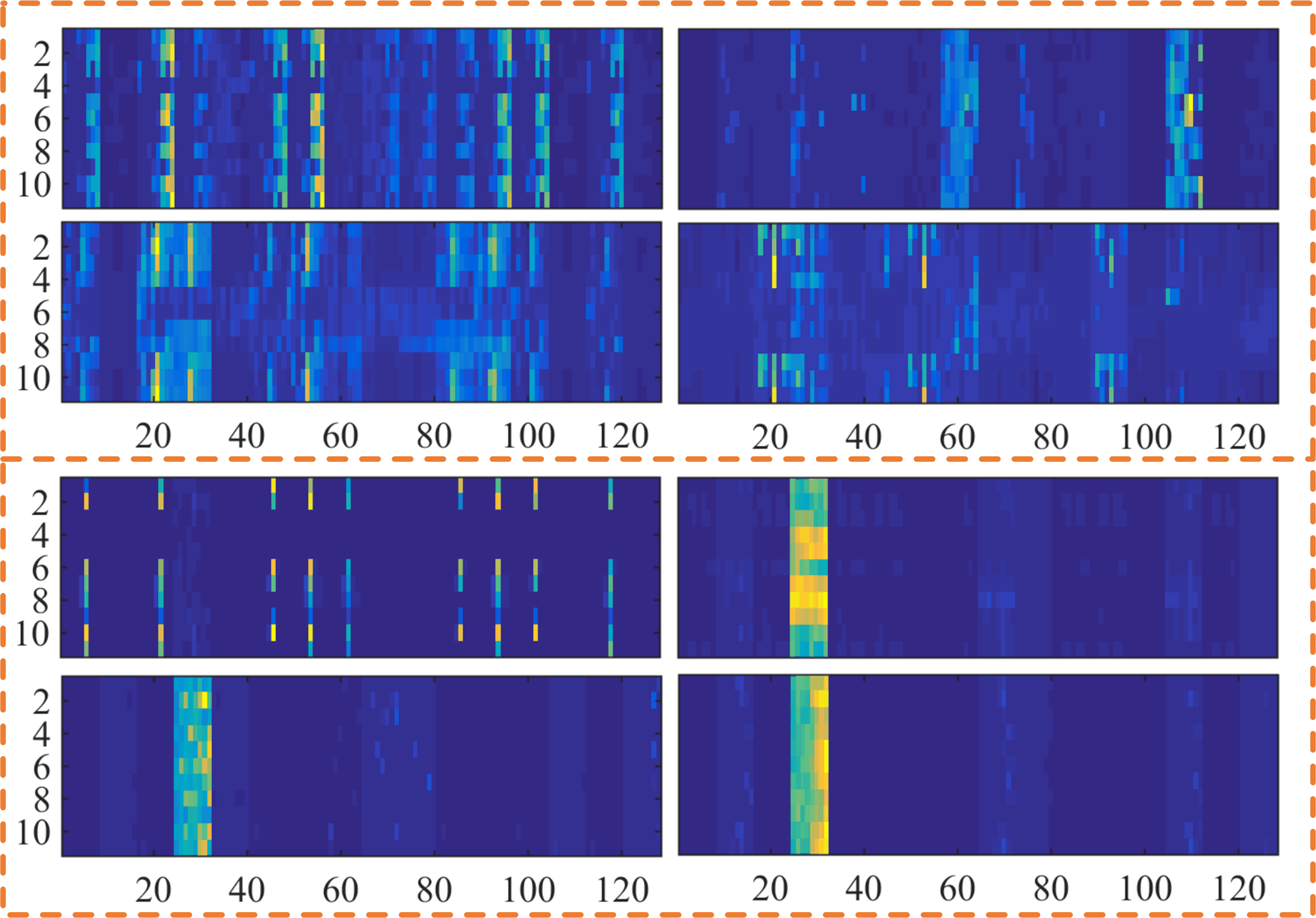}
  \caption{Visualization of four pairs of discriminative acoustic representations. Each row is corresponding to one audio clip. The top part of the box shows the examples when the model trained and test both on warblrb10k. Bottom part is the examples when model trained on BirdVox-20k and test on ff1010bird.}
  \label{fig:vis}
\end{figure}


In Fig.\ref{fig:vis}, we visualize four audio samples'
acoustic representations of the final layer of each discriminator.
The vertical axis represents the frame index, and the horizontal axis
represents the frequency index of all stacked channels.
The upper and lower parts within the dashed box respectively show
the representations on in-domain and cross-domain testing.
The 1st and 3rd rows indicate the audio representations
with bird calls while the 2nd and 4th rows refer to the ones without bird calls.
It can be observed that each pair in the 1st and 3rd rows are very different.
Each pair in the 2nd and 4th have something in common which represent the background sounds.
According to this visualization, we can conclude that the parallel discriminator is able to produce the discriminative representations as we expect.

\section{Conclusion}
\label{sec:con}


This paper investigates a new CNN-based architecture for acoustic 
event detection task to alleviate the domain mismatch problem, 
which features two CNN discriminators and an additional discriminative loss.
In addition, we design two kinds of training strategy and two alternative 
binary classifiers to further improve the system performances.
Experiment results on DCASE2018 task3 dataset have shown 
that our two-stage training strategy with frame-wise classifier 
significantly outperforms the baseline system in most cross-domain 
evaluation cases.


\bibliographystyle{IEEEtran}

\bibliography{template}

\end{document}